# Next Generation Opportunistic Networking in Beyond 5G Networks

Baldomero Coll-Perales[1], Loreto Pescosolido[2], Javier Gozalvez[1], Andrea Passarella[2], and Marco Conti[2]

[1] UWICORE laboratory, Universidad Miguel Hernández de Elche, Elche, Alicante, Spain

[2] Institute for Informatics and Telematics, Italian National Research Council, Pisa, Italy

bcoll@umh.es, loreto.pescosolido@iit.cnr.it, j.gozalvez@umh.es, a.passarella@iit.cnr.it, marco.conti@iit.cnr.it

*Abstract*— **Beyond 5G networks are expected to support massive traffic through decentralized solutions and advanced networking mechanisms. This paper aims at contributing towards this vision through the integration of device-centric wireless networks, including Device-to-Device (D2D) communications, and the Next Generation of Opportunistic networking (NGO). This integration offers multiple communication modes such as opportunistic cellular and opportunistic D2D-aided communications. Previous studies have demonstrated the potential and benefits of this integration in terms of energy efficiency, spectral efficiency and traffic offloading. We propose an integration of device-centric wireless networks and NGO that is not driven by a precise knowledge of the presence of the links. The proposed technique utilizes a novel concept of graph to model the evolution of the networking conditions and network connectivity. Uncertainties and future conditions are included in the proposed graph model through anticipatory mobile networking to estimate the transmission energy cost of the different communication modes. Based on these estimates, the devices schedule their transmissions using the most efficient communication mode. These decisions are later revisited in real-time using more precise knowledge about the network state. The conducted evaluation shows that the proposed technique significantly reduces the energy consumption (from 60% to 90% depending on the scenario) compared to traditional single-hop cellular communications and performs closely to an ideal "oracle based" system with full knowledge of present and future events. The transmission and computational overheads of the proposed technique show small impact on such energy gains.**

## I. INTRODUCTION

5G networks are mainly supported by the ultra-dense deployment of infrastructure-centric cellular solutions. However, in view of the increasing trend in traffic demand patterns [1], the sole scaling of the current one-hop infrastructure-centric cellular solutions will likely stress the network performance in terms of spectrum use and energy consumption. To address these issues, beyond 5G networks will pursue more decentralized and automated network paradigms [1][2]. This trend will increase the capabilities of future networks to anticipate to changes and predict the evolution of network connectivity and networking conditions. This anticipatory (mobile) networking paradigm represents a significant improvement with respect to traditional networking processes that are mostly designed to react once the changes have already happened [3].

In this context, the design of future networks will rely on advanced networking mechanisms and intelligent software that benefit from the use of data analytics and shared contexts and knowledge [1]. This vision of future networks foresees a further evolution that moves the edge of the network to the smart devices. This device-centric paradigm is aimed at exploiting the capabilities of devices such as smartphones, connected vehicles, machines, or robots [4]. In a long term view, these devices would become a more integral part of the network, by sharing their networking, computing, and sensing capabilities, just like other infrastructure resources [2]. This is achieved by the active participation of the devices in the operation of the networks through carefully designed cooperation and coordination mechanisms with the cellular infrastructure.

Device-centric wireless networks, including Device-to-Device (D2D) and Multi-hop Cellular or D2D-aided cellular communications, can utilize more efficiently the device's and network's resources when combined with opportunistic networking. Unlike traditional opportunistic networking mechanism that is used in disconnected networks, the Next Generation of Opportunistic networking (NGO) is not driven by the current presence and state of the links but by their potential to efficiently support the requested demand and services in a given time window [5], [6]. NGO is then aimed at exploiting the best connectivity opportunities. For example, NGO could schedule transmissions over an established link based on the channel state and benefit from the (long-term or short-term) predictive knowledge of this state to pause/resume the transmissions in order to improve their efficiency and reliability, and reduce the channel utilization and energy consumption [5]. In this context, NGO is particularly appealing to networks that do not suffer disconnections, like cellular networks, where connectivity is (normally) not an issue. The operation of NGO in cellular networks can also benefit from the known geographic context, including the location of the Base Station (BS) and spatial layout of buildings, to anticipate or predict the channel conditions based on the distance and visibility between the device and the BS.

The efficient integration of device-centric and NGO will further foster the role of devices as sources of information, and therefore will contribute to increasing the share of uplink on the mobile data traffic. An important aspect to consider is that, while the amount and size of contents generated from human users (e.g., in connection with the use of social networks, like instant messages, tweets, photos, videos, etc.) keeps increasing, the emerging role of devices as the *source* of automated data is also becoming a more and more important aspect to take into account in the design of the new networking paradigms. Both types of data (human generated and automated) can be related to a wide range of applications that often impose less stringent constraints in terms of latency with respect to the needs of real-time services[1]. This enables new degrees of freedom in the design of networking mechanisms. These degrees of freedom allow to pursue in a more effective way the optimal use of the energy resources in battery-powered devices. These important challenges bring an opportunity for the design of effective and energy efficient networking mechanisms that exploit device-centric and NGO.

In this work, we propose a novel networking technique that integrates device-centric and NGO. The proposed technique has been carefully designed to schedule the transmissions of delay-tolerant contents generated by mobile devices to the BS. Contents can be transmitted either directly or through D2D-aided communications to the BS. The transmission window is determined by the content's delay tolerance constraint. To schedule these transmissions, the proposed scheme exploits a novel concept of graph that is used to represent the evolution of the networking conditions and network connectivity, and where uncertainties and future conditions are included through anticipatory mobile networking. In particular, it considers stochastic geographic and link-context contextual information, such as the devices spatial density and distribution and spatial layout of buildings and streets, which can be made available in cellular networks at relatively low cost. Using this stochastic knowledge, the proposed scheme performs an offline estimate on the communication mode and time instant to perform the transmissions. A key feature of the proposed scheme is its real-time adaptation when more precise context information is available. In particular, the proposed scheme continuously revisits the selected transmission mode and scheduling decisions, and it updates them if a more efficient combination is found.

We demonstrate that the proposed technique achieves an efficient integration of NGO and device-centric wireless networks. The obtained simulation results show that the proposed technique reduces significantly (by up to 90% under the evaluated scenarios) the energy consumption in the data transmissions with respect to traditional single-hop cellular communications. The proposed technique also performs closely to an optimal scheme that also integrates NGO and device-centric wireless networks, but that assumes full knowledge of the future network state and can thus decide the optimal transmission schedule and communication mode. We have also analyzed the transmission and computation overheads required by the execution of the proposed technique. This analysis has shown that these overheads do not compromise the achieved energy gains.

The rest of this paper is organized as follows. Section II reviews related studies. Section III introduces the communication system and formulates the problem to be solved in this work. Section IV presents this paper proposal to efficiently integrate NGO and device-centric wireless networks. The proposed technique makes use of a novel graph to model the evolution of the network connectivity and networking conditions, which is described in Section V. Section VI shows how the energy cost of the different communication modes is estimated leveraging anticipatory knowledge. The practical implementation of the proposed technique is reported in Section VII using pseudocode, and its performance evaluation is provided in Section VIII. Finally, Section IX summarizes what are the main outcomes of this work and concludes the study.

## II. RELATED WORK

The trend towards device-centric wireless solutions has been fostered recently by the identified benefits of D2D and D2D-aided cellular communications, also known as Multi-hop Cellular Networks (MCN) or User Equipment (UE)-to-network relaying [7]. D2D-aided cellular communications allow (mobile) devices to connect to the cellular infrastructure through intermediate (mobile) devices, i.e. they integrate D2D and cellular communications. As it has been highlighted by 3GPP under Release 15, this is of particular interest to IoT (e.g. wearables) devices which have the benefit of almost always being in close proximity to a smartphone that can serve as a relay [7]. Device-centric wireless networks in general, and D2D-aided cellular communications in particular, are recognized as a core component of 5G (and Beyond) networks. Indeed, a recent 3GPP study item under Release 17 proposes to further explore device-centric solutions to improve the energy-efficiency and coverage of 5G scenarios and verticals such as in Home, Smart Farming, Smart Factories, Public Safety use cases, among others [8].

More in general, the idea of integrating D2D communication in the architecture of a cellular network dates back to the work of Lin and Hsu [9], where a MCN architecture was proposed. In that, and many subsequent works, the possibility to leverage delay tolerance and mobility aspects were not considered. Furthermore, the system setup relied on a random channel access at the MAC layer, based on RTS/CTS mechanisms. In the last decade, researchers have been working on the integration of D2D in wideband multicarrier-based cellular networks (i.e., 4G, 5G, and beyond). The initial studies focused on local (proximity) services (see, e.g., [10]). Lately, he research on D2D data offloading techniques, e.g. [11], has widened the scope of integrating D2D and opportunistic networking concepts, in order to investigate efficient ways to deliver contents obtained from remote servers. In this context, it has been showed that considerable energy savings can be achieved by exploiting content caching at the mobile devices, delay-tolerance, and the nodes mobility [6], [12], [13].

Opportunistic networking has traditionally focused on self-organized/ad-hoc mobile networks that lack end-to-end connections. In these scenarios, mobile devices can temporarily store the information, and eventually forward it to another device, which is more likely to be within the communication range of the destination, when a connection opportunity arises. Further gains can be obtained in cellular networks where devices tend to have a direct connection to the infrastructure almost everywhere thanks to the increasing densification of the infrastructure. Therefore, end-to-end connectivity is not an issue. In this context, the NGO paradigm would allow to integrate opportunistic networking into cellular networks to establish connections based on their efficiency and effectiveness and not just their presence. These benefits can be even higher when NGO is combined with D2D and D2D-aided (or MCN) solutions. For example, in [14] the authors consider a two-hop uplink communication where the intermediate mobile device, with store, carry and forward capabilities, relays the transmission between the source node and the BS. An optimization framework is proposed to identify the optimum locations at which the D2D and cellular transmissions should take place to minimize the energy consumption. Since it is not always feasible that the source node finds a mobile device at the derived optimum location and time instant, the authors propose in [6] a set of strategies (AREA and DELAY) that build from [14] and exploit context information, such as density and distribution of devices, to facilitate their implementation. In particular, in the AREA strategy the source node searches for devices to perform the D2D transmission in an area around the derived optimum location; this area guarantees with certain probability the presence of at least once device. The DELAY strategy estimates the time the D2D transmission should be delayed to guarantee with certain probability the presence of a device at the derived optimum location. The work presented in [15] derives the upper-bound capacity gains of the integration of NGO and device-centric wireless networks. To this aim, [15] proposes to use space-time graphs that are dynamic sequences or snapshots of the network topology representing the time evolution of nodes' locations and connectivity. Space-time graphs are used to model all possible D2D and cellular links along the time available to complete an upload. Then, an optimization framework is used to select the set of D2D and cellular links that minimize the cellular channel utilization, while guaranteeing that the upload is completed within the available time. These studies have revealed the potential and upper-bound benefits of integrating NGO and D2D-aided communications in cellular networks. However, most of these studies require a complete or deterministic knowledge of the nodes' location and their connectivity which is not realistic in practice.

---

[1] Non-real-time services (e.g. social networking, cloud services, data metering, mobile video, urban sensing, smart factories, smart farming, public safety use cases, etc.) will represent an important share of the forthcoming mobile data traffic, according to recent estimates [1].

The benefits of the integration of NGO and D2D-aided cellular networks have been also recently demonstrated empirically. In particular, the results reported in [5] show spectral efficiency gains up to a factor of 4.7 and 12 in outdoor pedestrian and vehicular scenarios, respectively. The tests included opportunistic D2D-aided cellular transmissions where the D2D and cellular links were pre-defined at the beginning of each trial.

The studies conducted to date have shown the benefits and the potential of integrating NGO and device-centric technologies into cellular networks, but they rely on deterministic models, assumptions or pre-defined links that prevent their implementation in real networks. A preliminary approach towards this integration was presented by the authors in [16]. This work significantly extends [16] by: (i) proposing a novel graph-based representation of the evolution of the network, (ii) providing the mathematical support for the computation of the expected cost of all the links based on anticipatory knowledge, (iii) introducing a two-phase procedure which is designed to revisit, in a second-phase that operates in real-time using more accurate context information, the long-term scheduling and mode selection decisions made offline, (iv) enabling intermediate devices to participate in the planning of when and how the information should be transmitted in a more active manner, without further involvement of the original source node. The resulting strategy exploits the reduction of uncertainty and the consequent use of more accurate context information to reduce the cost of the transmissions.

## III. COMMUNICATION SYSTEM AND PROBLEM FORMULATION

Without loss of generality, this study focuses on uplink transmissions where a source node $s$ needs to transmit a content of size $D_c$ to the BS before a deadline $T_{max}$. In the considered scenario, the use of NGO and device-centric wireless networks offers multiple communication mode options that can be selected to perform the transmission (Figure 1). The traffic delay tolerance also provides an additional degree of freedom in the selection of the best scheduling for transmitting the content. A sketch of the communication modes available in the considered scenario is provided in Figure 1. In traditional single-hop cellular communications the source node transmits directly the content to the BS at time instant $t_0$, when the content is generated (bottom-left quadrant of Figure 1). Opportunistic cellular communications can postpone the transmission to a later time, say $t_k$ (top-right quadrant of Figure 1). The decision of postponing the cellular transmission can be motivated, on the basis of some anticipatory context knowledge, by the estimation/prediction that the cellular transmission will be more efficient at $t_k$ than an immediate transmission at $t_0$. Finally, the integration of opportunistic networking and D2D-aided cellular communications provides an additional degree of freedom for exploiting intermediate mobile nodes and performing multi-hop transmissions to the BS[2] (top-left quadrant). In this case, the transmission times of all the hops in the opportunistic D2D-aided cellular communication can be scheduled across the available time $T_{max}$.

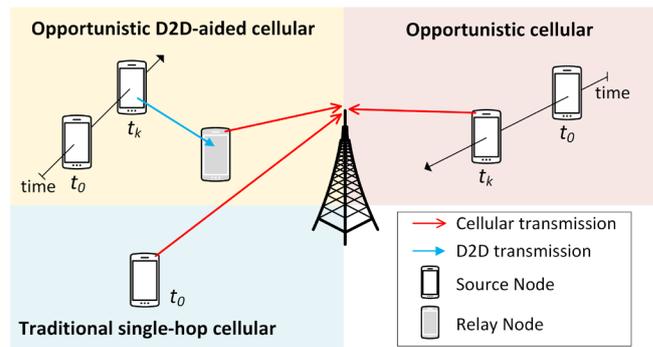

Figure 1: Traditional single-hop cellular, opportunistic cellular, and opportunistic D2D-aided cellular communication modes.

In the considered scenario, time is discretized using a step equal to $T_{CI}$ that is defined as the duration of a control interval (CI). A CI refers in this work to a time unit during which the network conditions may be considered to not change significantly. We consider that in each CI both D2D and cellular wireless links are allocated with an amount of resources (namely, Physical Resource Blocks (PRBs)) for transmitting a maximum of $D_{CI}$ bits. $D_{CI}$ bits represent a fragment or chunk of the larger content of size $D_c$ bits (the content is then divided in $N_c = \lceil D_c/D_{CI} \rceil$ fragments). This leads to the definition of $N_{CI}$, that is equal to $T_{max}/T_{CI}$, and that indicates the number of CIs available to complete the transmission of the $N_c$ fragments (note that in the considered scenario $N_c$ cannot be higher than $N_{CI}$). The time frame within which the transmission should be completed is then composed of the CIs starting at time instants $t_k = k \cdot T_{CI}, k \in \{0, 1, \ldots, N_{CI} - 1\}$, where $t_0$ indicates the starting time instant of the first CI successive to the arrival of the transmission request (in the source devices) by the application layer.

The defined scenario and communication system bring a large set of options over which decisions have to be made.

---

[2] The opportunistic D2D-aided cellular transmission represented in Figure 1 is limited to two-hops, but the multi-hop operation could be extended to integrate more intermediate devices.

First, for the transmission of each of the $N_c$ fragments, it has to be selected the communication mode among the ones illustrated in Figure 1. Then, to fully exploit the potential of NGO, each transmission can be scheduled at any time instant $t_k = k \cdot T_{CI}, k \in \{0, 1, \ldots, N_{CI} - 1\}$. It is important to note that the scheduled transmissions do not need to be consecutive in time and should take place when the best communication conditions are predicted. In general, the selection of the communication mode and transmission scheduling can target different performance metrics, e.g., energy consumption, spectrum efficiency, throughput, fairness, etc. Given the importance of energy efficiency for 5G networks and beyond, we focus in this work on minimizing the transmission energy consumption. In this context, all these decisions could be made offline based on an anticipatory context knowledge of the evolution of the network state and networking conditions. To realize these anticipatory context knowledge, this works relies on the support of the BS (see details in Section IV about the information the BS makes available to the mobile devices in the cell). However, the predicted or anticipated conditions might change as the transmission progresses, and the scheduling and communication mode decisions made offline could not be the ones that minimize the transmission energy consumption anymore. Therefore, the designed technique should be flexible enough to adapt to the changing network conditions.

## IV. D2D-AIDED NEXT GENERATION OPPORTUNISTIC NETWORKING

This work proposes a 2-phase technique to facilitate the integration of NGO and D2D-aided communications into cellular networks. The proposal is summarized in Figure 2. The proposed technique is designed to select both the communication mode and the time instant that are estimated to be the most efficient to complete the transmissions. These decisions are first made offline and based on anticipated or predictive knowledge of the evolution of the network state (see '1st phase: Planning' in Figure 2). Based on the premise that more accurate estimations can be made for short time windows [3], the proposed technique also includes a second phase (see '2nd phase: Execution' in Figure 2). This 2nd phase is executed in real time and is utilized to revisit the mode selection and scheduling decisions made in the 1st phase. The 2nd phase decisions are made using more precise knowledge about the network state.

Following the nomenclature introduced in Section III, the 2-phase technique proposed in this work is designed to derive the subset of $N_c$ control intervals within the time limit $T_{max}$ where fragments should be uploaded, and the communication mode to use for each of the $N_c$ required transmissions. And the aim of these selections is to minimize the transmission energy consumption.

The 1st phase, named *Planning*, takes as inputs the $N_c$ fragments of the content $D_c$. The *Planning* phase is in charge of estimating at $t_0$ the energy cost of transmitting a fragment using opportunistic cellular and opportunistic D2D-aided cellular transmissions for each CI of the time window available to complete the transmission (of duration $T_{max}$). To this aim, the *Planning* phase benefits from a novel concept of graph (Section V) that models all connection possibilities between the mobile nodes and the BS, and leverages the geographic and link context attributes of anticipatory mobile networking to estimate/predict the energy cost of these connections (Figure 2). The support from the BS is assumed in this *Planning* phase to make available such context attributes to the mobile nodes in the cell. Geographic and link context attributes refer to statistical models about the density and distribution of nodes within the cell, crossing probabilities at intersections, expected trajectories of the nodes, and contextual information in the form of street maps and *nominal* path loss maps. Section VIII.C discusses how the BS obtains and distributes this information in the cell, and what is the cost for transmitting it. Using this stochastic information, the *Planning* phase derives the communication modes (opportunistic single-hop cellular –'SH cellular' or opportunistic D2D-aided – 'D2D-aided' in Figure 2) that entail the lower energy cost for $N_c$ different CIs. For example, Figure 2 shows that as a result of the *Planning* phase a 'SH cellular' transmission is scheduled at time instant $t_0$, and 'D2D-aided' transmissions are scheduled at $t_3$ and $t_{N_{CI}-2}$. It is important to note that in this 1st phase the selection of the scheduling and communication modes is based on an *anticipated* probabilistic knowledge that might not be precise.

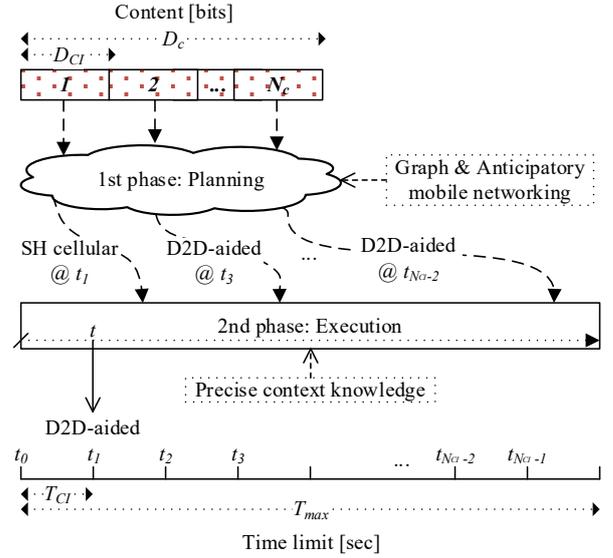

Figure 2: Representation of the proposed technique.

The 2nd phase, named *Execution*, takes as inputs the decisions made during the *Planning* phase, and revisits them to make the final scheduling and mode selection decisions. Contrary to the *Planning* phase that is executed at $t_0$ (when the content is generated), the *Execution* phase is performed in real time just before the start of each CI. Another difference with respect to the *Planning* phase is that the *Execution* phase benefits from more precise context knowledge rather than stochastic anticipatory knowledge. For example, the *Execution* phase relies, at each point in time, on the true link connectivity conditions measured by the mobile devices and the support of D2D discovery mechanisms to identify neighboring devices. The example illustrated in Figure 2 showed that, as a result of the *Planning* phase, a 'SH cellular' transmission was scheduled at $t_1$. However, based on the context information available in the *Execution* phase at $t_1$, a 'D2D-aided' cellular transmission is finally scheduled in this CI. The *Execution* phase could also decide to schedule a transmission in a CI that was not initially selected during the *Planning* phase. This could be the case if the *Execution* phase derives that a transmission (either 'SH cellular' or 'D2D-aided') in this CI is more efficient than any of the coming transmissions scheduled in the *Planning* phase.

Finally, the proposed technique is also designed to let all devices, not only the source and destination, contribute to the network operation. This is pursued in the proposed technique when an opportunistic D2D-aided cellular transmission is scheduled. In this case, the device selected to forward the information to the BS is empowered to take its own decisions on how (i.e. communication mode) and when (i.e. transmission scheduling) to forward the received fragment. These decisions are made following the two phases described above, adapted to the remaining time to complete the transmission, and considering that the content to transmit is just a single fragment. The selected device could then override the decisions of the source node if, through the context information it is aware of, it estimates that the network state and networking connectivity conditions are different.

## V. Graph Model

Graph theory is commonly utilized in opportunistic networking-related studies as it provides the necessary models and tools to characterize the dynamism of the network [17]. Existing types of graphs, such as wireless graphs, contact graphs, social graphs or time-varying graphs, utilize the graph's vertices and edges to represent the network's nodes and the presence of wireless links between nodes, respectively. For example, in wireless graphs, the presence of an edge indicates that two nodes are within the communication range of each other, and that the link between them can support a minimum data rate. These types of graphs are utilized for networks that suffer disconnections, and assuming that links are established as soon as nodes are within the communication range of each other. In addition, the creation of these graphs is only possible if it is assumed, for example, that the nodes' present and future locations and their connectivity are deterministic and known in advance. These sorts of assumptions are not realistic in practice though. Time-varying graphs (see [18][19]) introduce advanced tools for modeling dynamic systems where the network structure (i.e. graph's vertices and edges) vary over time. In particular, time-varying graphs utilize multilayer structures where edges connect nodes in the same or different layers depending on the temporal characteristic. However, these graphs also rely on finite and discrete knowledge of the network. In this context, this work proposes a novel graph representation of the evolution of the network connectivity where uncertainties and future conditions are taken into account through anticipatory mobile networking. In particular, the proposed graph exploits stochastics models and context information that can be made available in cellular networks, and leverages the use of new types of graph's vertices and edges.

### A. Graph based representation

Figure 3 illustrates an example of the graph proposed in this work to model the D2D-aided NGO. The vertices of the proposed graph can be expressed as $\mathcal{V} = (S \cup D \cup R)$, where $S$ is the source node $s$ willing to transmit a content, $D$ is the BS that is the destination of the content, and $R$ is an AREA/REGION that represents the D2D coverage area around $s$ and within which it will search for a relay that would forward the information to the BS. The AREA/REGION vertex $R$ is included in the proposed graph to account for the uncertainty (during the initial *Planning* phase) on the availability of devices that could act as candidate relays. It should be noted that $R$ does not represent a particular device, but an area/region where there might be any number of devices (or none). Therefore, a relay $r$ within this region/area needs to be selected if the $S$–$R$ edge is utilized.

Taking the vertex $S$ as example, the nomenclature we follow in the vertices is as follows: $S_{X,i}^k$ where $k$ indicates the index of the current CI (i.e. $k \in \{0, 1, \ldots, N_{CI} - 1\}$), $i$ indicates the "birth" instant or time instant the vertex first appears in the graph (i.e. $i \in \{0, 1, \ldots, N_{CI} - 1\}$), and $X$ is a vector that includes the subscript of the vertex's parent and shows the family tree; the length of $X$ also shows the number of hops the fragment has traveled so far. For example, $S_{0,1}^2$ represents a source node $S$ at time instant $t_2$, that received a fragment at time instant $t_1$, and whose parent is the original source node (i.e. subscript 0). The AREA/REGION vertex $R$ follows the same nomenclature. The vertex $D$, that represents the BS, does not include any sub-/superscript as it keeps the same properties (e.g. same location) across time.

The proposed graph includes two different types of edges: intra-graph's edges and inter-graphs' edges (see Figure 3). The intra-graph's edges represent the wireless links between $S$–$D$, $S$–$R$ and $R$–$D$. These edges are expressed as $\mathcal{E}_a = \{(S, D), (S, R), (R, D) \mid S, D, R \in \mathcal{V}\}$. The inter-graphs' edges are referred to as $\mathcal{E}_e$ and they are made of the edges $\gamma$ and $\delta$, i.e. $\mathcal{E}_e = \gamma \cup \delta$. The inter-graphs' edges $\gamma$ connect the same source node $S$ at successive CIs, i.e. $\gamma_{X,i}^k = \left(S_{X,i}^k, S_{X,i}^{k+1}\right)$. The inter-graphs' edges $\delta$ connect the REGION/AREA vertex $R$ of a source node $S$ at the $k$-th CI, with another vertex $S$ at the $(k + 1)$-th CI, i.e. $\delta_{X,i}^k = \left(R_{X,i}^k, S_{(X,i),k+1}^{k+1}\right)$. Then, the inter-graphs' edges set is $\mathcal{E}_e = \{\left(S_{X,i}^k, S_{X,i}^{k+1}\right), \left(R_{X,i}^k, S_{(X,i),k+1}^{k+1}\right) \mid S, R \in \mathcal{V}; \, k, i \in \{0, 1, \ldots, N_{CI} - 1\}$. Note that the superscript and subscript of the inter-graph's edges borrow the nomenclature of the vertices $S$ and $R$. Overall, the set of edges of the proposed graph are expressed as $\mathcal{E} = \mathcal{E}_a \cup \mathcal{E}_e$.

An important feature of the proposed graph's vertices is the transformation from an AREA/REGION vertex $R$ at a certain instant $t_k$ into a source node vertex $S$ at instant $t_{k+1}, k \in \{0, 1, \ldots, N_{CI} - 2\}$. This stands for the proposed empowerment of the selected relay to takes its own decisions with data fragment(s) it has just received. In this context, the relay selected within the AREA/REGION $R$ during the $k$-th CI becomes a new vertex $S$ from instant $t_{k+1}$ onwards. The new vertex $S$ is in charge of transmitting the received fragment to the BS (vertex $D$) either by itself through a direct cellular, or through another relay that will eventually be present in the (new) AREA/REGION $R$ from time instant $t_{k+1}$ onwards. This is represented in Figure 3 by the sub-graphs aligned in the right column. It should be noted that, since the relay that would be selected within the AREA/REGION $R$ at the $k$-th CI is not known in advance, these sub-graphs are created on the fly as the transmission progresses.

In this context, the purpose of the inter-graphs' edges $\mathcal{E}_e$ is as follows. The inter-graph's edges $\gamma$ are used to report the source node at successive CIs the transmission progress (i.e. remaining content and time to complete the transmission). The inter-graphs' edges $\delta$ connect a REGION/AREA vertex $R$ with the selected relay $r$ located within such area. The selected relay $r$ becomes a vertex $S$ at the next CI. In this context, inter-graphs' edges $\delta$ are used to report the selected relay the fragment to be transmitted and the time available to complete the transmission.

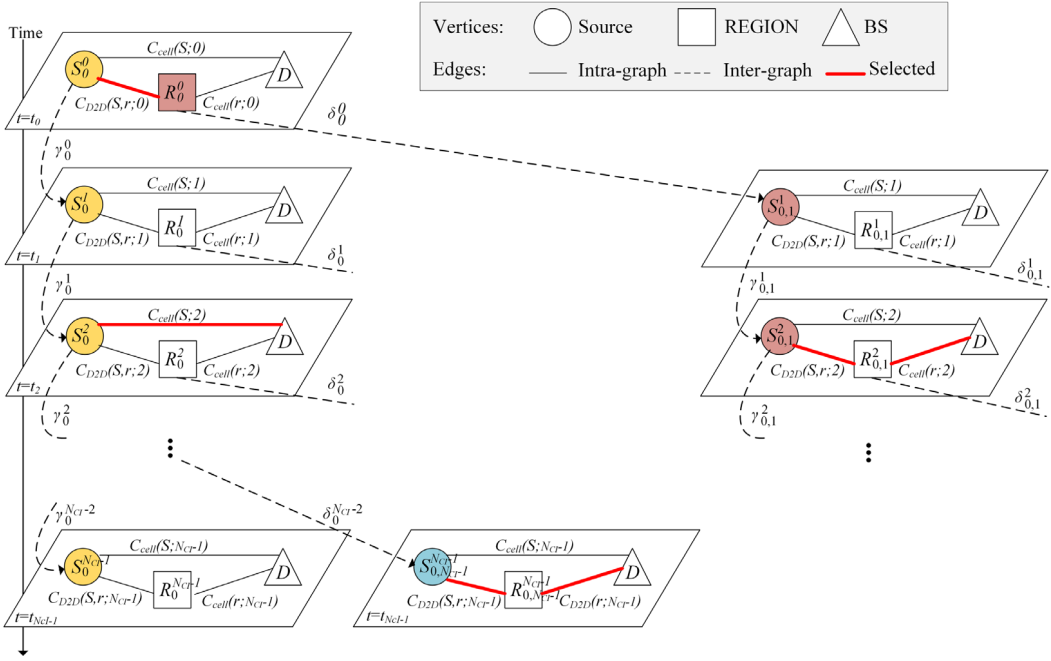

Figure 3: Graph-based representation of the next generation opportunistic networking concept.

Finally, the weights of the intra-graph's edges use the notation $C(x; k), k \in \{0, 1, \ldots, N_{CI} - 1\}$, to indicate the cost of the cellular link from either the vertex $S$ ($C_{\text{cell}}(S; k)$), or a relay $r$ selected within the REGION/AREA $R$ ($C_{\text{cell}}(r; k)$), to the BS, and the cost of the D2D link from the source $S$ to the relay $r$ ($C_{\text{D2D}}(S, r; k)$). Further details on how these costs are computed are provided in Section VI. For the sake of notation simplicity, we have dropped the vertices the dependence on super-/subscripts.

### B. Communication modes modeled by the proposed graph

Following the definitions provided in the previous subsection, the proposed graph models all possible communication modes between the source node $s$ and the BS:

i. Single-hop (SH) traditional: a direct (immediate or non-deferred) cellular transmission from $s$ to BS (see in Figure 3 the link between $S_0^0$ and $D$ at time instant $t_0$);

ii. Opportunistic cellular: a direct but deferred SH transmission between $S$ and BS (see in Figure 3 the link between $S_0^k$ and $D$ at $t_k$, $\forall k \in \{1, \ldots, N_{CI} - 1\}$);

iii. D2D-aided cellular: a multi-hop cellular transmission from $s$ to the BS that includes a D2D transmission from $s$ to a relay $r$ found in the AREA/REGION $R$, and a cellular transmission from $r$ to BS (see in Figure 3 the links between $S_0^0$ and $R_0^0$, and between $R_0^0$ and $D$ at $t_0$);

iv. Opportunistic D2D-aided cellular: a deferred D2D transmission from $s$ to a relay $r$ found in the AREA/REGION $R$ that forwards the content to the BS (see in Figure 3 the links between $S_0^k$, $R_0^k$, and $D$ at $t_k$, $\forall k \in \{1, \ldots, N_{CI} - 1\}$);

v. Opportunistic multi-hop D2D-aided cellular: an immediate or deferred D2D transmission to a relay $r$ found in the AREA/REGION $R$, which becomes a vertex $S$ and performs any of the previous communications modes to transmit the received fragment (see in Figure 3 the graphs hanging from the inter-graphs' edges $\delta_0^k$, $\forall k \in \{0, 1, \ldots, N_{CI} - 2\}$. Theoretically, the opportunistic multi-hop D2D-aided operation is not constrained to a number of hops. The graphs that would hang from $\delta_{0,1}^1$ and $\delta_{0,1}^2$ in Figure 3 represent these multi-hop scenarios.

Out of all these possibilities, the source node $s$ independently selects the communication mode, and schedules the time instant at which the transmissions should take place to minimize the transmission cost. It should be noted that the source node can only take these decisions on the graphs that are connected through the inter-graphs' edges $\gamma$. When the source node selects a (opportunistic) D2D-aided communication mode, it delegates the fragment transmission to the selected relay (inter-graph's edge $\delta$). Then, the selected relay can choose the time instant and communication mode to use for the fragment transmission. In the example illustrated in Figure 3, the source node $S_0^k, k \in \{0, \ldots, N_{CI} - 1\}$, schedules a D2D-aided cellular transmission at time instant $t_0$, and a direct cellular transmission at $t_2$. For the D2D-aided cellular transmission, the source node $S_0^0$ grants the selected relay in the AREA/REGION $R_0^0$ the freedom to select the communication mode (i.e. either -deferred- cellular or D2D-aided) to use for the transmission of the received fragment. In order to make such decision, the selected relay computes the cost *estimates* of the opportunistic cellular and D2D-aided cellular transmissions from $t_0$ onwards. For the example illustrated in Figure 3, the relay selected at $t_0$, which becomes $S_{0,1}^1$, schedules an opportunistic D2D-aided cellular at $t_2$. Finally, at $t_2$, the selected relay in $R_{0,1}^2$ estimates that forwarding the fragment to the BS is the communication mode that minimizes the transmission cost.

## VI. Energy Cost Estimation

This work exploits the *geographic and link context attributes* of anticipatory mobile networking to make offline estimates of the intra-graph's edges cost. In particular, for each $S_{X,i}^k$, it is estimated the energy cost of performing an opportunistic single-hop cellular transmission at the k-th CI, which we indicate with $C_{cell}(S; k), \forall k \in$ $\{0, 1, \ldots, N_{CI} - 1\}$), and an opportunistic D2D-aided cellular transmission, which includes the cost of the D2D transmission, $C_{D2D}(S, r; k)$, and the relay-to-BS transmission, $C_{cell}(r; k), \quad \forall \ k \in \{0, 1, \ldots, N_{CI} - 1\}$). Without loss of generality, this study sets the cost of the intra-graphs' edges to account for the energy cost of transmitting a fragment of size $D_{CI}$ bits.

Based on geographical context, the potential prediction of a user mobility can be as high as 93%, and the space-time mobility can be predicted using statistical methods [3]. This is exploited in this work that considers that each source node knows its own expected trajectory for an amount of time equal to the time limit $T_{max}$. When the expected trajectory reaches an intersection corner, this work accounts on turning probabilities since the source node might follow any direction at this critical location. With regards to all other nodes in the cell, we consider that the BS makes available to the devices in the cell the information about the density of mobile devices present in the cell, and their distribution. This also happens with a local map of the region, representing the dimension and size of the streets and buildings (Section VIII analyzes the cost of transmitting this information).

On the other hand, the anticipatory mobile networking's link context attribute is used to estimate the evolution of the physical wireless channel so that it is possible to take advantage of future link improvements or to counteract bad conditions before they impact the transmissions [3]. The link context attribute is exploited in this work in terms of a nominal path loss map that is maintained by the BS and made available to the devices in the cell (Section VIII analyzes the cost of transmitting this information which is made available to the devices periodically –cellular path loss map– and on-demand –D2D path loss map). This refers to the cellular path loss between any location in the area and the BS, and the D2D path loss within the D2D communication range. We indicate the cellular path loss with $g_{cell}(x, x_b)$, where $x$ is the location of the cellular transmitter, $x_b$ is the location of the BS, and the D2D path loss with $g_{D2D}(x, y)$, where $x$ is the position of the D2D transmitter and $y$ the position of the D2D receiver. Basically, the nominal path loss depends on the geometry of the system, and not on the instantaneous channel conditions that are subject to time varying effects such as shadowing, fast fading and frequency selectivity. Therefore, the path loss maps change over large periods of time. In particular, we consider that the nominal path loss maps $g_{cell}(x, x_b)$ and $g_{D2D}(x, y)$ are computed offline, or acquired through measurements sent regularly by the devices and suitably processed, over a very large time scale, by the network operator. Random channel effects, such as multi-path and shadowing, are hard to predict and this work consider them unknown. To account for these random channel effects, a link margin $M$ is added to the path loss upon computing the transmit power of each transmission.

### A. Opportunistic single-hop cellular transmissions

We indicate with $\hat{C}_{cell}(S; k)$ and $\hat{C}_{cell}(r; k)$, $\forall \ k \in \{0, \ldots, N_{CI-1}\}$, the expected energy cost of transmitting a fragment of $D_{CI}$ bits from the source node $s$ (or graph vertex $S$) or a relay $r$, respectively, to the BS through a cellular link at time instant $t_k$. Taking as example the cellular transmission from the source node $s$, the expected energy cost of the single-hop cellular transmission can be computed as

$$\hat{C}_{cell}(S; k) = C_{cell}(x_s(t_k), x_b), \tag{1}$$

where $C_{cell}(x_s(t_k), x_b)$ is the energy cost of transmitting a fragment from the source node located at $x_s$ at the $k - th$ CI and $x_b$ is the location of the BS. The (discrete-time) expected location of a source node $S_{X,i}^k$ at the $k - th$ CI can be represented as $x_s(t_k) = x_s(t_0) + v_s \cdot k \cdot T_{CI}$, where both the location ($x_s$) and speed ($v_s$) vectors belong to $\mathbb{R}^2$. $C_{cell}(x_s, x_b)$ is a function of the cellular path loss introduced earlier ($g_{cell}(x, x_b)$), and of the cellular technology utilized to perform the transmission. In general, the transmit power ($P_{tx}$) required to guarantee that the signal power level at the receiver is equal to a threshold $P_{tx}^{Th}$ can be computed as $P_{tx} = P_{tx}^{Th}/g$. Then, the energy cost can be computed as a function of $P_{tx}$, i.e. $C = f(P_{tx})$. The specific relation $f$: $P_{tx} \rightarrow C$ depends on the considered communication technologies (Section VIII.A shows how the energy cost is computed for the considered evaluation scenario). It should be noted that $C_{cell}(x_s(t_k), x_b)$ is a deterministic quantity. $\hat{C}_{cell}(S; k)$ could then be estimated at any control interval k, $\forall \ k \in \{0, \ldots, N_{CI} - 1\}$. For the estimation of $\hat{C}_{cell}(S; k)$, this work relaxes the need to know the expected trajectory of a source node beyond a critical location like an intersection corner. In particular, if it is considered that the source node $s$ reaches an intersection corner $I_i$ at time instant $t_i$, the expected energy cost of the single-hop cellular transmission at time instant $t_{i+1}$ can be computed as

$$\hat{C}_{cell}(S; i + 1) = C_{cell}(x_s^r(t_{i+1}), x_b) \cdot P_r^i \tag{2}$$
$$+ C_{cell}(x_s^l(t_{i+1}), x_b) \cdot P_l^i$$
$$+ C_{cell}(x_s^f(t_{i+1}), x_b) \cdot P_f^i,$$

where $P_r^i$, $P_l^i$ and $P_f^i$ are the turning probabilities for right, left and forward directions at the intersection $I_i$ and $x_s^r(t_{i+1})$, $x_s^l(t_{i+1})$ and $x_s^f(t_{i+1})$ are the respective expected locations of the source node at time instant $t_{i+1}$ if it takes any of these directions, respectively. The expression in (2) could be generalized to trajectories that include more than one turning event.

### B. Opportunistic D2D-aided transmissions

We indicate with $\hat{C}_{D2D-aided*}(S; k)$ the expected energy cost of an opportunistic D2D-aided transmission at the $k - th$ control interval. $\hat{C}_{D2D-aided*}(S; k)$ includes the cost of both transmissions, i.e. the D2D transmission from the source node to the relay, and the cellular transmission from the relay to the BS. The asterisk in $\hat{C}_{D2D-aided*}(S; k)$ indicates that the source node will eventually select, among the available potential relays, the relay that minimizes the overall energy consumption of the D2D-aided transmission.

The estimation of the energy cost $\hat{C}_{D2D-aided*}(S; k)$ is computed taking into account that the locations of the relays are unknown. Instead, stochastic information about their spatial density and distribution within the cell is exploited. We indicate with $R_s(t_k)$ the nominal D2D coverage region of a source node $s$ at given time instant $t_k$. This coverage region is defined as a disk of radius $R_{D2D}$ centered at $x_s(t_k)$, deprived of unreachable spaces for D2D transmissions like, for instance, locations inside buildings and streets. $R_s(t_k)$ is represented as a tessellation of square tiles of equal surface (e.g. 1m²). Let $1, \ldots, Q$ be an arbitrary labeling of the tiles available at $R_s(t_k)$, and $x_i(t_k), \forall i = 1, \ldots, Q$, be the center of the tiles at time instant $t_k$. The total energy cost of transmitting a fragment using a D2D-aided cellular transmission from a source node $s$ located at $x_s$ in $t_k$, and using a relay $r$ located at the center of tile $i$ of the coverage region $R_s(t_k)$, i.e., at $x_i(t_k)$, can be computed as

$$
\begin{aligned}
C_{\text{D2D-aided}}(S;k) = \; & C_{\text{D2D}}\big(x_s(t_k), x_i(t_k)\big) \\
& + C_{\text{cell}}(x_i(t_k), x_b),
\end{aligned} \tag{3}
$$

where $C_{\text{D2D}}(x, y)$ is the energy cost of the D2D transmission from the location $x$ to the location $y$, and $C_{\text{cell}}(x, x_b)$ is the energy cost of the cellular transmission (see Section VI.A).

The costs $C_{\text{D2D-aided}}(S;k)$, computed for each tile $i \in \{1, \dots, Q\}$, allow to establish a ranking of the locations (i.e., the tiles) in which it would be more preferable to have a relay. For a given position $q$ in the ranking, we indicate with $i(q)$ the labeling index of the tile at position $q$ in the ranking. Conversely, for a given tile $i$, we indicate with $q(i)$ the position of the tile $i$ in the ranking. All possible values of the energy cost that would be incurred by the D2D-aided cellular transmission at $t_k$ can be expressed as $c_1(t_k), \dots, c_Q(t_k)$, where the numbering order follows the energy cost-based ranking order, i.e., $c_1(t_k) \leq c_2(t_k) \leq \dots \leq c_Q(t_k)$.

Consider a relay whose position, within $R_s(t_k)$, is distributed according to some probability density function $p(x)$. We approximate this continuous bidimensional distribution with a discrete random variable whose possible outcomes are the tiles in which the relay may fall. We indicate the discrete probability distribution of the relay position with $p(i)$. Then, a given relay position falls in the tile $i$ of the tessellation of $R_s(t_k)$ with probability $p(i)$. The probability distribution of the relay's position over the tiles, $p(i)$, induces a probability distribution of the energy cost that would be associated to the D2D-aided cellular transmission. We indicate this probability distribution with $p_C(c_q)$, where, for the sake of notation simplicity, we have dropped the dependence of $c_q$ on the time instant $t_k$. It should be noted that $p_C(c_q) = p(i(q))$. In this context, the Cumulative Distribution Function (CDF) of the energy cost of the D2D-aided cellular transmission can be expressed as the following staircase function:

$$
\begin{aligned}
F_{C_{\text{D2D-aided}}}(c) &= \sum_{q=1}^{Q} p_C(c_q)\, u(c - c_q) \\
&= \sum_{q=1}^{Q-1} \left( \sum_{q'=1}^{q} p_C(c_{q'}) \right) \cdot u_{[c_q, c_{q+1}]}(c) + u(c - c_Q),
\end{aligned} \tag{4}
$$

where $u_{[a,b]}(c)$ is the unit rectangular function (equal to 1 if $c \in [a, b]$ and 0 otherwise), and $u(\cdot)$ is the unit step function.

To compute the expected energy cost of the D2D-aided cellular transmission, which is used by our proposed system as an input to the scheduling algorithm, we use an approach similar to that exploited, in a different context[3], in [12]. Let us assume that at time instant $t_k$ there are $J$ relays in the D2D coverage region $R_s(t_k)$ of the source node $s$. The $J$ relays are numbered as $r_1, \dots, r_J$. The relays' positions are statistically independent and identically distributed according to the same probability distribution $p(i)$. The statistical independence of the relays entails statistical independence of the energy cost that would be incurred by using any of them. Therefore, the energy costs of the $J$ relays are identically distributed random variables with a common CDF equal to (4). Assume that, out of the $J$ relays, the relay with the lowest energy cost, which we indicate as "best relay", is selected. Then, the cost of the D2D-aided cellular transmission is the minimum of the $J$ above referenced i.i.d. random variables. We indicate this energy cost as $\widehat{C}_{\text{D2D-aided}*}(S; k)$. The CDF of this energy cost can be computed as:

$$
F_{\widehat{C}_{\text{D2D-aided}*}}(c \mid J) = 1 - \left(1 - F_{C_{\text{D2D-aided}}}(c)\right)^J, \tag{5}
$$

and the corresponding discrete probability distribution of the lowest energy cost for the D2D-aided cellular transmission as:

$$
p_C(c_q \mid J) = \tag{6}
$$
$$
= \begin{cases}
1 - \left(1 - p_C(c_q)\right)^J & \text{if } q = 1 \\
1 - \left(1 - \sum_{q'=1}^{q} p_C(c_{q'})\right)^J - \sum_{q'=1}^{q-1} p_C(c_{q'} \mid J) & \text{if } q \in \{2, \dots, Q-1\} \\
1 - \sum_{q'=1}^{Q-1} p_C(c_{q'} \mid J) & \text{if } q = Q.
\end{cases}
$$

It should be noted that to compute (5) and (6) it was assumed that the number of relays $J$ within the D2D coverage region is known, but this is, in reality, an unknown random number that it is estimated utilizing the statistical information about the nodes density and distribution within the cell. To compute the statistics of the number of relays that will be located within the D2D coverage region at time $t_k$, we model the presence of the nodes in each street as a unidimensional Spatial Poisson Point Process (SPPP). More specifically, we label the streets in the entire cell with the numbers in the set $\Psi \triangleq \{1, \dots, N_\Psi\}$ and indicate with $\lambda_\psi$ the (linear) density of nodes present on street $\psi \in \Psi$. Then, $\sum_{\psi=1}^{N_\Psi} \lambda_\psi$ corresponds to the spatial density of nodes in the cell. For each time instant $t_k$, we consider the subset $\Psi_{k^{(s)}} \triangleq \{\psi_1^{(s,k)}, \dots, \psi_{N}^{(s,k)}\} \subset \Psi$ of the streets whose median axis is at least partially within the coverage region $R_s(t_k)$, and we also indicate with $l_n^{(s,k)}$ the length of the portion of street $\psi_n^{(s,k)}, \forall n \in \{1, \dots, N^{(s,k)}\}$, covered by $R_s(t_k)$. We indicate with $J_n^{(s,k)}, \forall n \in \{1, \dots, N^{(s,k)}\}$, the random variable representing the number of relays in the segment of the street $\psi_n^{(s,k)}$, and with $J^{(s,k)} \triangleq \sum_{n=1}^{N^{(s,k)}} J_n^{(s,k)}$ the overall number of relays within $R_s(t_k)$. By construction, $J_n^{(s,k)}$ is a Poisson random variable with average value $\mathbb{E}(J_n^{(s,k)}) = \lambda_{\psi_n^{(s,k)}} l_n^{(s,k)}$. The overall number of nodes within the coverage region $R_s(t_k)$, in the considered model, is the sum of the statistically independent variables of this kind, and therefore is itself a Poisson random variable with mean

$$
\mathbb{E}(J^{(s,k)}) = \sum_{n=1}^{N^{(s,k)}} \mathbb{E}(J_n^{(s,k)}) = \sum_{n=1}^{N^{(s,k)}} \lambda_{\psi_n^{(s,k)}} l_n^{(s,k)}, \tag{7}
$$

and distributed as:

$$
p_{J^{(s,k)}}(j) = \frac{\mathbb{E}(J^{(s,k)})}{j!} \exp\left(\mathbb{E}(J^{(s,k)})\right). \tag{8}
$$

The distribution of the energy cost of the D2D-aided cellular transmission from the source node s at time instant $t_k$ can be expressed as:

$$
F_{\text{D2D-aided}*}^{(s,k)}(c) = \sum_{j=0}^{\infty} F_{\text{D2D-aided}*}^{(s,k)}\big(c \mid J^{(s,k)}\big) p_{J^{(s,k)}}(j), \tag{9}
$$

where, recalling (4) and (5),

---

[3] In [12] the objective is to compute the best candidate to offload packets transmission in the downlink direction by nodes that have them cached, i.e., considering a single hop. Here, the overall problem is complicated by the need to take into account two or more hops (BS-to-Relay and Relay-to-Destination).

$$F_{\text{D2D-aided}^*}^{(s,k)}\left(c\middle|J^{(s,k)}\right) = \sum_{q=1}^{Q-1}\left(1-\left(1-\sum_{q'=1}^{q}p_c\left(c_{q'}^{(s,k)}\right)\right)^{J^{(s,k)}}\right) \quad (10)$$
$$\cdot\, u_{\left[c_q^{(s,k)},c_{q+1}^{(s,k)}\right]}(c) + u\left(c-c_Q^{(s,k)}\right).$$

The expected energy cost of the D2D-aided cellular transmission can be computed as:

$$\hat{C}_{\text{D2D-aided}^*}^{(s,k)}\left(e\middle|J^{(s,k)}\right) = \sum_{j=0}^{\infty}\left(p_{J^{(s,k)}}(j)\sum_{q=1}^{Q}p_{C^*}\left(c_q^{(s,k)}\middle|j\right)c_q^{(s,k)}\right) \quad (11)$$
$$= \sum_{q=1}^{Q}\left(c_q^{(s,k)}\sum_{j=0}^{\infty}\left(p_{J^{(s,k)}}(j)p_{C^*}\left(c_q^{(s,k)}\middle|j\right)\right)\right)$$

In practice, (11) can be approximated by limiting the infinite sum over $j$ (i.e. number of relays). The considered values for $j$ can be selected according to an interval $([j_{le}(s,k), j_{ue}(s,k)])$ in which, for instance, 95% of the probability mass of the corresponding distribution (6) is concentrated. Then, the expected energy cost of the D2D-aided cellular transmission from the source node $s$ at time instant $t_k$ can be computed as:

$$\hat{C}_{\text{D2D-aided}^*}(S;k) = E_{\hat{C}_{\text{D2D-aided}^*}^{(s,k)}}\left(c\middle|J^{(s,k)}\right) \quad (12)$$
$$= \sum_{q=1}^{Q}\left(c_q^{(s,k)}\sum_{j=j_{le}^{(s,k)}}^{j_{ue}^{(s,k)}}\left(p_{J^{(s,k)}}(j)p_{C^*}\left(c_q^{(s,k)}\middle|j\right)\right)\right).$$

Again, $\hat{C}_{\text{D2D-aided}^*}(S;k)$ could be estimated at any control interval $k$, $\forall\ k\in\{0,\dots,N_{CI}-1\}$ relying on the expected trajectory of the source node . As we did in (2) for the estimation of the energy cost of opportunistic single-hop cellular transmissions, the estimation of $\hat{C}_{\text{D2D-aided}^*}(S;k)$ relaxes when the expected trajectory of the source node $s$ reaches the intersection corner $I_i$ at time instant $t_i$. In this case, the expected energy cost of the opportunistic D2D-aided cellular transmission at time instant $t_{i+1}$ can be computed as:

$$\hat{C}_{\text{D2D-aided}^*}(S;i+1) = \hat{C}_{\text{D2D-aided}^*}(S^r;k)\cdot P_r^i \quad (13)$$
$$+\ \hat{C}_{\text{D2D-aided}^*}(S^l;k)\cdot P_l^i$$
$$+\ \hat{C}_{\text{D2D-aided}^*}(S^f;k)\cdot P_f^i,$$

where $P_r^i$, $P_l^i$ and $P_f^i$ are the turning probabilities for right, left and forward directions at the intersection $I_i$, and $S^r$, $S^l$, $S^f$ are expected locations of the source node at time instant $t_{i+1}$ if takes any of these directions, respectively, i.e. $x_S^r(t_{i+1})$, $x_S^l(t_{i+1})$ and $x_S^f(t_{i+1})$. (13) could be generalized to trajectories that include more than one turning event.

## VII. Implementation of NG Opportunistic Networking

This section describes through pseudocode a practical implementation of the proposed technique introduced in Section IV, that exploits the knowledge generated using the graph presented in Section V and the use of anticipatory mobile networking to estimate the cost of the transmissions (Section VI). The proposed algorithm is carried out in two phases named *Planning*, and *Execution*. The second phase involves two different steps that we have named *real-time adaptation*, and the actual *transmission*. First, this section formally defines the problem formulation introduced in Section III.

### A. *Algorithm setup variables*

In the considered scenario, the source node $s$ needs to transmit a content of size $D_C$ before the time limit $T_{\max}$. The content is divided in $N_c = [D_C/D_{CI}]$ fragments. The time is organized in CI of duration $T_{CI}$, so that the source node $s$ has $N_{CI}$ ($N_{CI} = T_{\max}/T_{CI}$) control intervals to complete the transmission. In the considered scenario $N_c < N_{CI}$. The initialization and definition of the scenario's traffic variables is represented in **Pseudocode 1**. 1:6. In this context, the goal of the proposed algorithm is to find the subset of $N_c$ control intervals where fragments should be uploaded, and the communication mode to use for each of the $N_c$ required transmissions, in order to minimize the transmission energy consumption.

**Pseudocode 1:** Algorithm configuration

1. //*Source node $s$ has $D_c$ bits to transmit before $T_{\max}$*
2. **Set** $T_{CI}$ to the duration of a control interval (CI)
3. **Set** $D_{CI}$ to the content size to be transmitted in a CI
4. //*s* defines the scenario's traffic variables
5. $N_c = \text{ceil}(D_C/D_{CI})$      // frags. that need to be transmitted
6. $N_{CI} = T_{\max}/T_{CI}$    // control intervals to transmit the frags.

### B. *Planning* phase

This phase is executed by the source node $s$ as soon as the content to be uploaded is generated. This phase leverages the knowledge generated from the graph modeling presented in Section V to select the CIs and communication modes to use for the transmission of each $N_c$ fragments.

In particular, the selection of CIs and communication modes is based on the estimates of the energy cost of performing opportunistic cellular ($\hat{C}_{\text{cell}}$) and opportunistic D2D-aided cellular transmissions ($\hat{C}_{\text{D2D-aided}^*}$) across the available $N_{CI}$ control intervals (see Section VI). As explained in Section VI, these estimates are obtained exploiting geographic and link-context contextual information made available in the cellular networks. Using the estimates of the energy cost for each of the communication modes across the $N_{CI}$ control intervals, $s$ schedules the transmission of each fragment in a control interval and decides what communication mode to use for its transmission. To do so, it follows this energy cost ranking procedure: it first selects at each control interval the communication mode incurring a lower energy cost (**Pseudocode 2.** 10:14); 2) the resulting energy cost estimates are ranked in ascending order (**Pseudocode 2.** 15:17); 3) then, the first $N_c$ estimates, out of the $N_{CI}$ computed, with minimum energy cost are selected (**Pseudocode 2.** 18:20). As a result of the *Planning* phase, the *selected_CI* and *selected_mode* variables are set with the time instants and communication modes of the $N_c$ first ranked energy cost estimates. These variables indicate the strategy computed offline for the transmission of the content from the source node.

**Pseudocode 2:** *Planning phase*

7. **Initialize** set of *selected_CI* $[N_c] = \{\ \}$
8. **Initialize** set of *selected_mode* $[N_c] = \{\ \}$
9. //Estimate energy cost of communication modes
10. $\hat{C}_{\text{cell}}(S;k) \to f(\text{context info.}), \forall\ k\in\{0,1,\dots,\ N_{CI}-1\}$
11. $\hat{C}_{\text{D2D-aided}^*}(S;k) \to f(\text{context info.}), \forall\ k\in\{0,1,\dots,\ N_{CI}-1\}$
12. //Select at each control interval the communication mode

13. //with lower energy cost
14. $\hat{C}(S; k) = \min\left(\hat{C}_{cell}(S; k), \hat{C}_{D2D-aided*}(S; k)\right), \forall k \in \{0, 1, \ldots, N_{CI} - 1\}$,
15. //Rank $\hat{C}(S; k)$ in ascending order keeping track of
16. //indexes of control intervals
17. $[C_{min}, Index] = sort(\hat{C}(S; k), \text{'ascending'})$
18. //Select the first $N_c$ elements
19. $selected\_CI = Index(1: N_c)$
20. $selected\_mode = mode(C_{min}(1: N_c))$

### C. Execution

#### 1) Real-time adaptation

The outputs of the *Planning* phase are the *selected_CI* and *selected_mode* variables that indicate the time instants at which the fragment transmissions should take place and the communication modes to use for such transmissions. This transmission strategy is computed at the start of the process (as soon as the content is generated) and is based on energy cost estimates across the $N_{CI}$ control intervals. On the other hand, the *real-time adaptation* step of the *Execution* phase is carried out in real-time just before (e.g. $\sigma$ seconds before, $\sigma \ll T_{CI}$) the start of the CIs. This step uses more precise knowledge about the network state (e.g. true link quality conditions measured at their current locations, rather than statistical path loss maps) to possibly revisit the scheduling and communication mode selection decisions made during the *Planning* phase. It should be noted that the *real-time adaptation* step acts both in the control intervals at which a transmission has been scheduled and in those that are free. Then, for each control interval, the *real-time adaptation* evaluates the cost of the opportunistic cellular and opportunistic D2D-aided cellular transmissions considering the real network conditions (**Pseudocode 3. 23:27**). For example, for the evaluation of the energy cost of the opportunistic SH cellular transmissions, the uncertainty of the expected trajectory of the source node is removed. For the evaluation of the energy cost of the opportunistic D2D-aided transmissions, besides the nodes density and distribution uncertainty, the uncertainties about the number of relays within the D2D coverage region (i.e. equations (5) and (6)) and about their link quality conditions with the source node and the BS, are removed. This is the case since during the *Execution* phase the nodes rely on standard neighbor discovery mechanisms (e.g. 3GPP's UE-to-Network neighbor discovery [20][21])[4]. This provides more accurate values of the energy costs than the ones estimated in the *Planning* phase.

Then, for the control intervals at which a transmission is scheduled, the communication mode that shows a lower energy cost calculated in the *real-time adaptation* step is selected. If this mode does not match the scheduled one in the *Planning* phase, then the *selected_mode* variable is updated (**Pseudocode 3. 29:35**). For the control intervals that are not scheduled, the *real-time adaptation* updates the scheduling decisions made in the *Planning* phase if the energy cost computed for the current CI is lower than any of the energy costs estimated for the coming scheduled transmissions. If this is the case, the update involves the cancellation of the coming scheduled transmission, and the execution of the derived transmission at the current time instant (**Pseudocode 3. 36:46**).

---

**Pseudocode 3:** *Real-time adaptation*

---

21. //Check just before ($\sigma$ *secs*) the start of the *CI*
22. **For** $t = t_k - \sigma, \forall k \in \{0, 1, \ldots, N_{CI} - 1\}$
23. //Estimate energy cost of communication modes at $t_k$
24. $\hat{C}_{cell}(S; k) \rightarrow f(\text{real context info.})$
25. $\hat{C}_{D2D-aided}(S; k) \rightarrow f(\text{real context info.})$
26. //Get the mode with minimum cost at $t_k$
27. $\hat{C}_{real-time}(S; k) = \min\left(\hat{C}_{cell}(S; k), \hat{C}_{D2D-aided*}(S; k)\right)$
28. //Any transmission scheduled at $t_k$?
29. **If** any($selected\_CI == t_k$)
30.   **If** $mode_k \mathrel{!}= selected\_mode_k$
31.     **If** $\hat{C}_{real-time}(S; k) < \hat{C}(S; k)$
32.     //Substitute the scheduled transmission at $t_k$
33.     $selected\_mode_k = mode(\hat{C}_{real-time}(S; k))$
34.     **End if**
35.   **End if**
36. **Else**
37.   **If** $\hat{C}_{real-time}(S; k) < \max(\hat{C}(S; i), i \geq k)$
38.   //Remove the scheduled transmission at $t_i$
39.   remove($selected\_CI, t_i$)
40.   remove($selected\_mode, mode\left(\hat{C}(S; i)\right)$)
41.   //Add the transmission at $t_k$
42.   add($selected\_CI, t_k$)
43.   add($selected\_mode, mode(\hat{C}_{real-time}(S; k))$)
44.   **End if**
45. **End if**
46. **End for**

---

#### 2) Transmission

Finally, the *transmission* step is in charge of implementing the communication modes selected in the previous *real-time adaptation* step at the scheduled control intervals. In the control intervals at which a direct cellular transmission is scheduled, the source node $s$ transmits directly the fragment to the BS (**Pseudocode 4. 53:54**). In the control intervals for which an opportunistic D2D-aided cellular transmission is scheduled, the source node selects a relay among its neighbors (the one entailing the lowest overall D2D-aided cellular transmission) that takes the responsibility of forwarding the fragment to the BS (**Pseudocode 4. 55:59**).

For the purpose of the technique proposed in this work, it could be considered that the selected relay becomes the source node of the received fragment with updated constraints: the content to be uploaded is just a fragment of size $D_{CI}$ and the time available to complete the transmission has been reduced from the original deadline ($T_{max}$). In this context, the selected relay performs the *Planning* phase, and later the *Execution* phase, to decide the communication mode to use for the transmission of this specific fragment. The relay might decide to forward the fragment itself to the BS as soon as received, defer the cellular transmission for some CIs, or to select another relay to forward the information to the BS. It should establishment where they would not reply to other source nodes' *requests* of a relay to connect to the BS.

---

[4] Based on the UE-to-Network neighbor discovery mechanism, source nodes and relays that have been already selected prevent being selected as relays for other transmissions. This is achieved during the neighbor discovery

be noted that the recursive process of forwarding a fragment to a relay could be considered endless if it is not limited by a system's hop count constraint. However, it is important to recall that the overall content needs to be uploaded before a deadline ($T_{max}$) which is inherited from the fragments. This will in turn result in a limit on the **number of hops/forwarding** that each fragment might undergo (**Pseudocode 4.** 60:61).

---

**Pseudocode 4:** *Transmission*

---

47. //Execute the selected communication mode at the
48. //scheduled *CI*
49. **For** $t_k, \forall\ k \in \{0, 1, \ldots,\ N_{CI} - 1\}$
50.   **If** any(*selected_CI* == $t_k$)
51.     **Switch** *selected_mode*($t_k$)
52.       //Direct transmission to the BS
53.       **case** *direct_cellular*
54.        *s* transmits the fragment to the BS
55.       //D2D-aided cellular transmission: the source node
56.       //selects a relay which takes responsibility of
57.       //transmitting this fragment with updated conditions
58.       **case** *D2D-aided_cellular*
59.        *s* selects a neighbor *r* that minimizes the energy cost of the D2D-aided cellular transmission
60.        *r* becomes *s'*
61.        **Goto Pseudocode 1** ($D'_c = D_{CI}$, $T'_{max} = T_{max} - t_k$)
62.     **End switch**
63.   **End if**
64. **End for**

---

## VIII. PERFORMANCE EVALUATION

### A. Evaluation environment

The performance of the proposed technique has been evaluated in a scenario of 6x6 blocks of a "Manhattan" grid. The main simulation parameters are summarized in Table I. The width of the buildings is 90m and the distance between sidewalks is 10m (6m-width streets are considered). The buildings have on average 4 floors of 5m height each. The BS is located at the center of the scenario at a height of 25m on top of a building. All this scenario layout and building characteristics are taken into account in the considered map-based channel model that is utilized to compute the nominal path loss for the cellular transmissions [22]. On the other hand, the METIS's deliverable D1.4 on channel models, which distinguishes between line-of-sight (LOS) and Non-LOS (NLOS) conditions, is utilized to model the nominal path loss for the D2D transmissions [23].

This work uses the SUMO (Simulation of Urban MObility) simulator to model the mobility of the nodes in the scenario [24]. SUMO has been configured to limit the speed of nodes to 1.5m/s, and to set equal probabilities at intersection corners of turning right or left, or continuing straight. We have considered scenarios with different densities of nodes ranging from 0.012 nodes/m to 0.18 nodes/m. This is equivalent to consider scenarios with a number of nodes ranging from 100 to 1500. For a scenario of this size, the simulation guidelines reported in [25] suggest considering 1500 nodes for the test case "dense urban scenario societies".

In the considered scenario, the source nodes are selected randomly among the available nodes, and the start of their transmissions follows a Poisson distribution with a rate $\lambda_{req} = 1/10$ (i.e. one source node is selected every 10s on average). The selected source node is requested to upload a content of size $D_c = 24$ Mbits which is fragmented in 6 fragments, i.e. $D_{CI}$

= 4 Mbits. The time limit to complete the transmission is set to $T_{max}$ {10, 20, and 30}s.

Without loss of generality, we consider that the cellular and D2D technologies share the same LTE spectrum band at 2.3GHz (a.k.a. in-band D2D), and the duration of the control interval $T_{CI}$ is set to 1s. Based on [26], each Physical Resource Block (PRB) carries a number of bits in the range from 16 to

TABLE I. SIMULATION PARAMETERS

| Symbol | Meaning | Value |
|---|---|---|
| *Propagation model* | | |
| $h_{BS}$ | Height of the BS | 25 m |
| $h$ | Avg. height of the buildings | 20 m |
| $n_{floors}$ | Avg. number of floors in the buildings | 4 |
| $h_{floor}$ | Height of floors | $h/n_{floors}$ |
| $Ws$ | Width of streets | 10 m |
| $Wb$ | Width of buildings | 90 m |
| $MCL$ | Minimum coupling losses | 70 dB |
| $M\_cell$ | Cellular link margin | 4 dB |
| $M\_D2D$ | D2D link margin | 10 dB |
| *D2D and cellular technologies* | | |
| $\tau_{PRB}$ | Duration of a Physical Resource Block (PRB) | 0.5 ms |
| $T_{CI}$ | Control Interval duration | 1s |
| $B_{PRB}$ | Bandwidth of a PRB | 180 KHz |
| $N_0$ | Noise power spectral density | -174 dBm/Hz |
| $R_{D2D}$ | Maximum D2D transmission range | 80 m |
| $n_u$ | Number of PRBs used to transmit a content fragment | 10.000 |
| *Traffic* | | |
| $D_c$ | Content size | 24M bits |
| $T_{max}$ | Time limit to complete the upload | {10, 20, 30}s |
| $\lambda_{req}$ | Content upload request rate | 1/10 reqs/s |

720 in LTE. Assuming that a PRB carries 400 bits, there would be needed 10.000 PRBs to transmit a content fragment of size $D_{CI}$ bits (i.e. 4Mbits). In the considered control interval of 1s, there are approximately {50.000, 100.000, 200.000, 500.000} PRBs for an LTE system of {5, 10, 20, 50} MHz bandwidth. In our simulations, we used a system bandwidth of 10 MHz This shows the cellular system can provide the required PRBs to transmit the $D_{CI}$ bits in a CI. How the cellular system deals with the management of radio resources within each slot is out of the scope of this work.

Considering the conditions described above for the D2D and cellular technologies, their transmission energy consumption is computed as follows. The relation between the (nominal) received power $P_{rx}$ and transmit power $P_{tx}$ is $P_{rx} = g \cdot P_{tx}$, where $g$ represents the nominal path loss. Let $n_u$ be the number of PRBs used to transmit the $D_{CI}$ bits within a control interval. The nominal Signal to Noise Ratio (*SNR*) at the receiver, associated to a PRB of bandwidth $B_{PRB}$ Hz can then be expressed as $SNR = P_{rx}/(N_0 \cdot B_{PRB}) = g \cdot P_{tx} / (N_0 \cdot B_{PRB})$, where $N_0$ is the noise power spectral density. The nominal capacity of the channel corresponding to the PRB, assuming no interference is $C = B_{PRB} \cdot log_2(1 + SNR) = B_{PRB} \cdot log_2(1 + g P_{tx} / (N_0 \cdot B_{PRB}))$. Then, the amount of information ($I_{CI}$) that can be transmitted using the allocated $n_u$ PRBs within a control interval is $I_{CI} = n_u \tau_{PRB} B_{PRB} \ log_2 (1 + g P_{tx} / (N_0 \cdot B_{PRB}))$, where $\tau_{PRB}$ is the duration of a LTE's slot, or PRB, and it is equal to 0.5ms, and $B_{PRB}$ is the bandwidth of a LTE's PRB equal to 180 KHz. In nominal conditions, the inequality $I_{CI} \geq D_{CI}$, which results in $P_{tx} \geq (1/g) \cdot \mathcal{N}_0 \cdot B_{PRB} \cdot \left(2^{(D_{CI}/(n_u \cdot \tau_{PRB} B_{PRB}))} - 1\right)$ would guarantee that the achievable amount of information transferred over the channel using the $n_u$ assigned PRBs is larger than the fragment size. Setting the transmit power to satisfy this inequality, however, would be enough in the absence of random fading and shadowing effects. To cope with such effects, we assume that the transmitter uses a link margin $M$ which guarantees that the *actual* amount of information that can be transferred exceeds $D_{CI}$ with a very high probability (e.g., 99%)[5]. The actual transmit power per PRB is hence set as $P_{tx} = M \cdot (1/g) \mathcal{N}_0 B_{PRB} \left(2^{(D_{CI}/(n_u \tau_{PRB} B_{PRB}))} - 1\right)$. In this context, the energy consumed in a control interval to transmit $D_{CI}$ bits can be computed as $E_{CI} = n_u \tau_{PRB} P_{tx}$. For the purposes of this work, we have computed, through offline simulations, suitable link margins for both cellular (4 dB) and D2D transmissions (10 dB), and used these values to set the transmit power.

### B. Configurations of the proposed technique and benchmarking schemes

We have analyzed the proposed technique under the following configurations in order to assess the impact of the envisioned phases (see Section VII) on the obtained performance:

- *Planning only*. This configuration implements the communication modes derived using the *Planning* phase at the scheduled time instants. If an opportunistic cellular transmission is scheduled, the source node directly transmits the fragment to the BS at the scheduled CI. When a D2D-aided cellular transmission is scheduled, the selected relay forwards the received fragment to the BS in the same CI. The *Execution* phase is limited to the *transmission* step; the *real time adaptation* step is not performed.
- *Planning + limited Execution (max 2 hops)*. This configuration performs the *Planning* phase and the *real-time adaptation* step of the *Execution* phase. However, when a D2D-aided cellular transmission is finally scheduled in the *real-time adaptation* step, the selected relay has to transmit the received fragment in the same CI (it cannot defer its transmission or forward the fragment to another relay).
- *Planning + complete Execution*. This configuration implements the complete two-phase technique introduced in Section VII. For the sake of practical feasibility, the multi-hop operation has been limited to the use of 2 relays (max 3 hops).

Besides, the following schemes have been also evaluated for benchmarking:

- Single-hop (SH) traditional. Source nodes implementing this scheme use the first $N_c$ control intervals to upload the content to the BS, i.e. opportunistic networking schemes are not implemented.
- Opportunistic cellular. For a fair comparison with the proposed scheme, the source nodes implementing the opportunistic cellular scheme use the *Planning* phase presented in Section VII (**Pseudocode 1** and **2**) to schedule the transmissions that minimize the energy consumption. However, the available communication modes are limited to opportunistic cellular transmissions. Then, only the estimates of the $\hat{C}_{cell}(S; k), \forall\, k \in \{0, \ldots, N_{CI} - 1\}$, are taken into account to schedule the opportunistic cellular transmissions.
- 5G-Relay. Source nodes implementing this scheme use the first $N_c$ control intervals to upload the content to the BS. If possible, the source nodes perform D2D-aided transmissions in the $N_c$ control intervals. For a fair comparison, the relay is selected following the procedure shown in the *real-time adaptation* step of the *Execution* phase (**Pseudocode 3.** 25). The source nodes use direct cellular transmissions in the CI where D2D-aided is not possible (e.g. no relays available in the D2D range).
- Optimum. Source nodes implementing this scheme have full knowledge about the network conditions and nodes' trajectories along the time. Then, source nodes can select the most energy-efficient communication modes (either opportunistic cellular or opportunistic D2D-aided cellular) for the $N_c$ control intervals (out of the $N_{CI}$ available ones) that minimize the energy consumption. The implementation of this scheme is unfeasible in real networks and it is used in this work to identify what is the upper-bound of opportunistic D2D-aided communications. Note that under this scheme the opportunistic D2D-aided cellular transmission is limited to 1 relay (i.e. 2 hops).

### C. Performance results

The results reported below are average values obtained over 100 simulation runs of 500 seconds of simulation time to guarantee the statistical accuracy of the results.

#### 1) Communication mode selection and scheduling

First, this section investigates how the different configurations of the proposed technique described above adapt to the context conditions of the scenario to select the communication modes and the control intervals to perform the fragment transmissions. Figure 5 shows the average ratio of the selected communication modes for the transmission of the content's fragments as a function of the number of nodes in the scenario. For example, Figure 5.a shows that the configuration including the sole *Planning* phase (i.e., with no real-time adaptation) selects the opportunistic SH cellular communication mode 39% (61% opportunistic D2D-aided cellular) of the times when there are 100 nodes in the scenario and $T_{max}$ is set to 10s. This percentage reduces to 2% (increases to 98%) when there are 1500 nodes in the scenario. The *Planning* phase captures the density of nodes in the scenario when it computes the estimates of the energy cost for the opportunistic SH cellular and opportunistic D2D-aided cellular transmissions. Increasing the number of nodes results in an increased likelihood that a relay is found in the D2D coverage region of the source node, and that the relays are at locations where the D2D-aided transmission requires a lower energy compared to the SH cellular transmission. Then, the opportunistic D2D-aided mode is selected more frequently as the density of nodes in the scenario increases.

The second configuration, that implements a limited *real-time adaptation* step of the *Execution* phase, revisits the expected performance of the communication modes at each control interval. This is performed using more precise knowledge about the network state (e.g. source node and relays' location uncertainties are removed) than in the initial *Planning* phase that performs these ("longer term") estimates offline for all the control intervals and then with higher uncertainties. The results reported in Figure 5.a show that when the source nodes implement the second configuration, they tend to use more frequently the opportunistic D2D-aided communication mode than when they implement the first configuration. Removing the nodes trajectories and location uncertainties results in a better assessment of when the opportunistic D2D-aided mode outperforms the opportunistic SH cellular communication mode. It is important to recall that decisions made during the *real-time adaptation* step happen both at control intervals for which the initial *Planning* phase scheduled a transmission, but also for those that are free. For the former case, the benefit is obvious since the conducted re-computation would help to correct a wrong estimation on the communication mode to use at the scheduled control interval. For the latter case, it can be decided to use a control interval that is free because it is computed that the energy cost incurred in this control interval by any of the communication modes is lower than the *energy cost estimate* for any of the coming scheduled control intervals. However, the *real energy cost* at that control interval is unknown at this point in time, and then the control interval swap might not be always beneficial. Figure 4 shows the main consequence of the real-time operation of the *Planning + limited Execution* configuration, that is the higher utilization of the first control intervals.

---

[5] Cellular networks define transmission modes to guarantee that the transmission error probability is below a certain threshold. This threshold can be as low as 0.00001 [27].

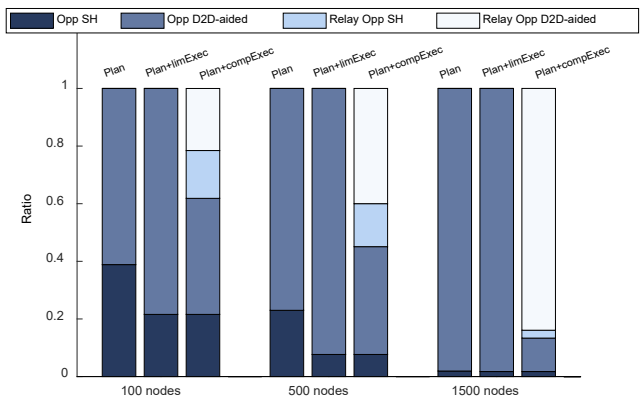

a) $T_{max}$ = 10s

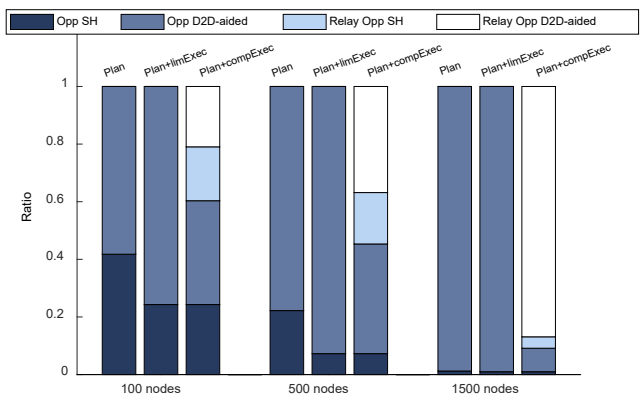

b) $T_{max}$ = 30s

Figure 4: Selected communication mode share for the *Planning only* ('Plan'), *Planning + limited Execution* ('Plan+limExec') and *Planning + complete Execution* ('Plan+compExec') configurations of the proposed technique.

Figure 4 shows a discrete probability function of the selected control intervals to perform the transmissions. The results reported in Figure 4.a for $T_{max}$ = 10s indicate that 86% (resp. 75%) of the transmissions performed when the source nodes implement this configuration use the first 6 CIs when there are 100 (resp. 1500) nodes in the scenario. This percentage reduces to 66% (resp. 60%) when the source nodes implement the *Planning only* configuration.

Finally, Figure 5.a also shows the communication modes selected when the source nodes implement the complete 2-phase technique proposed in this work (i.e. the *Planning + complete Execution* configuration). Under this configuration, when an opportunistic D2D-aided cellular transmission is scheduled, the selected relay can decide whether to forward the fragment to the BS immediately ('Opp. D2D-aided' in Figure 5) or to defer the cellular transmission ('Relay Opp SH' in Figure 5), or whether to select another relay ('Relay Opp. D2D-aided' in Figure 5) to forward the fragment. This decision is made by the selected relay utilizing, as first phase, the *Planning* phase of the proposed technique. Therefore, this decision should show similar trends to those already analyzed for the *Planning only* configuration, i.e. the selection of the 'Relay Opp. D2D-aided' communication mode should increase with the increasing number of nodes in the scenario. Indeed, Figure 5.a shows that when there are 100 nodes in the scenario, the selected relay uses the 'Relay Opp. D2D-aided' communication mode 21% of the times, and this percentage increases to 87% when the number of nodes in the scenario is 1500. The use of the *Planning + complete Execution* configuration results in changes in the control intervals selected to perform the transmissions (Figure 4). While the *Planning + limited Execution* configuration tends to use the firsts control interval because of the real-time reevaluation of the communication mode conditions, performing the *complete real-time adaptation* induces 30% (53%) of the uploads to utilize the last control interval available for the transmission of a fragment when there are 100 nodes (1500 nodes) in the scenario. This is the case because the *complete real-time adaptation* is able to select a relay that can fully exploit the

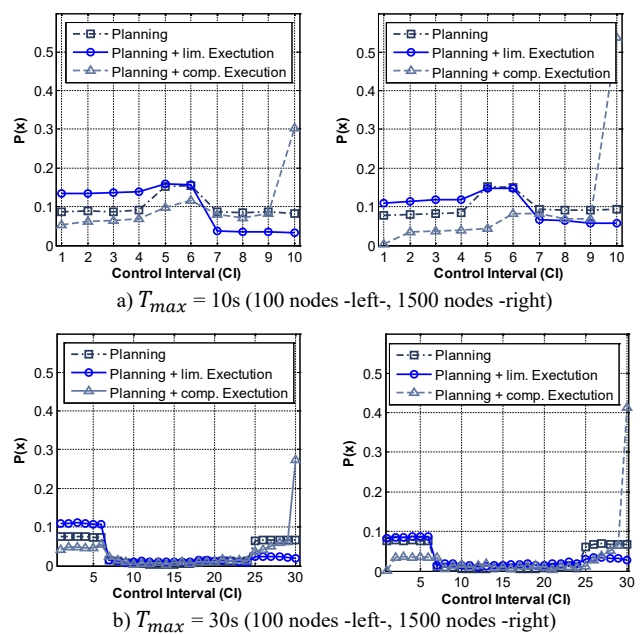

a) $T_{max}$ = 10s (100 nodes -left-, 1500 nodes -right)

b) $T_{max}$ = 30s (100 nodes -left-, 1500 nodes -right)

Figure 5: Distribution of control interval selection.

time available to find better conditions to communicate with the BS.

Figures are also reported for the scenarios with the time limit set to $T_{max}$ = 30s (see Figure 5.b and Figure 4.b) and show similar trends to those analyzed above.

### 2) Energy consumption for data transmission

This section benchmarks the energy consumption of the SH traditional, opportunistic cellular, 5G-Relay, and optimum schemes against that obtained with the different configurations of the proposed technique (see Figure 6). It should be noted that, out of all the schemes under evaluation, SH traditional is the only one that does not implement opportunistic networking or D2D-aided transmissions. Hence, Figure 6 uses the average energy consumption of SH traditional as a reference, over which the reduction (in percentage) of the average energy consumption achieved with the rest of schemes under evaluation is reported. The 95% confidence intervals of the obtained results are also depicted in Figure 6. The results reported in Figure 6 include the energy consumption of all nodes that participate in the uplink data transmission. In general, the obtained results clearly show that the use of opportunistic networking and D2D-aided transmissions help reducing the energy consumption compared to the traditional SH cellular communication. These benefits increase with the number of nodes in the scenario and the time available to complete the transmission. For example, the opportunistic cellular scheme reduces the energy consumption compared to SH traditional by 5.5% when $T_{max}$ is set to 10s (Figure 6.a), and by more than 20% when $T_{max}$ is set to 30s (Figure 6.c). In the opportunistic cellular scheme, the longer time window over which the content transmission can be completed allows the source nodes to delay the upload and schedule the fragments transmissions to those control intervals with a lower energy cost estimate.

The results reported in Figure 6 for the 5G-Relay scheme demonstrate that the only use of D2D-aided communications also helps reducing the energy consumption compared to the traditional SH cellular communication. In this case, the energy reduction levels increase with the number of nodes in the scenario as it allows the 5G-Relay scheme to better select a relay to perform an efficient D2D-aided transmission. For example, 5G-Relay reduces by 45% and 72% the energy consumption compared to SH traditional when the number of nodes in the scenario is 100 and 1500, respectively. It should be noted that 5G-Relay does not implement opportunistic networking (i.e. the transmissions are performed in the first $N_c$ control intervals). Therefore, 5G-Relay benefits do not depend on the time available to complete the transmission.

The energy consumption benefits compared to SH traditional increase when opportunistic networking is adequately combined with D2D-aided communications, as it is proposed in the technique presented in this work. The *Planning only* configuration of the proposed technique reduces by more than 50% and 65% the energy consumption compared to SH traditional when there are 100 nodes in the scenario and $T_{max}$ is set to 10s and 30s, respectively. The energy reduction levels go above 80% in both cases when the number of nodes in the scenario increases to 1500. While opportunistic networking allows exploiting the time dimension over which fragment transmissions can be scheduled, the integration with D2D communications brings a spatial dimension that can be utilized to find relays with more efficient links (located under better communication conditions) to the BS. The possibility to find such a relay increases with the number of nodes in the scenario. Indeed, the results reported in Figure 5 showed that the *Planning only* configuration tends to use the opportunistic D2D-aided communication mode with the increasing number of nodes in the scenario.

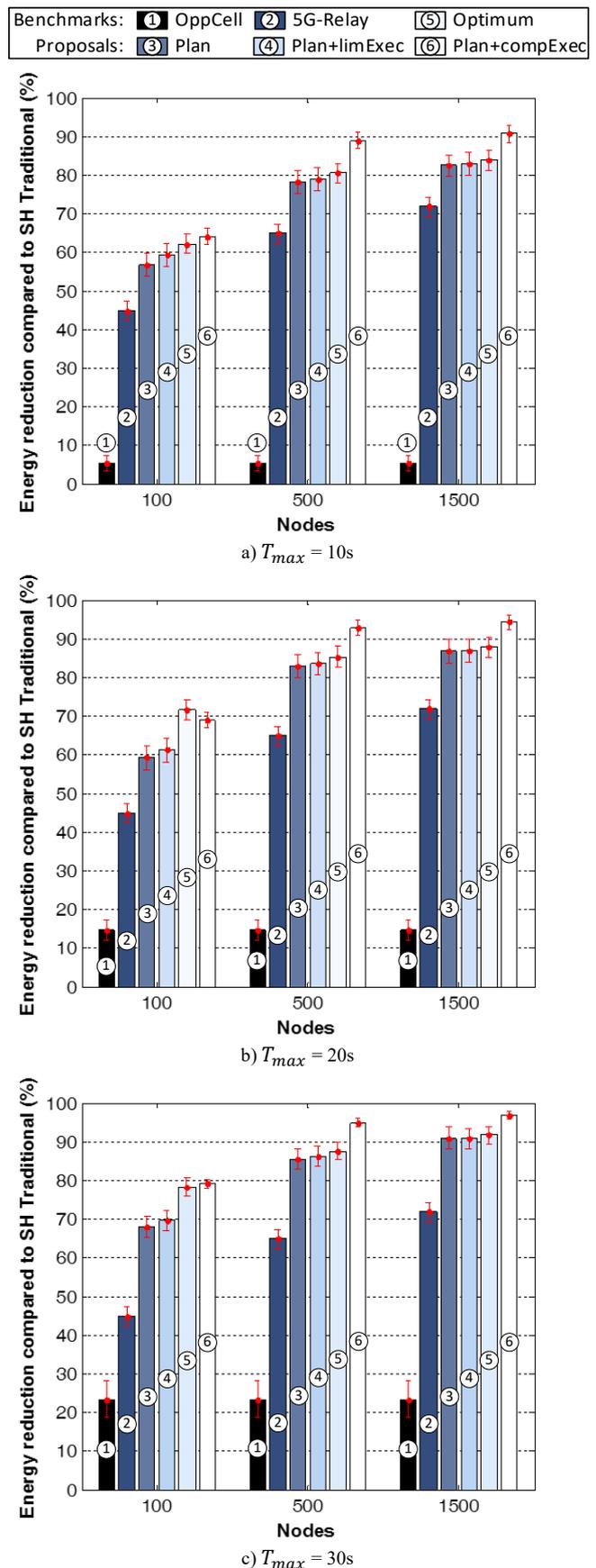

Figure 6: Reduction of energy consumption (in %) with respect to SH traditional for the opportunistic cellular ('OppCell'), 5G-Relay, *Planning only* ('Plan'), *Planning + limited Execution* ('Plan+limExec'), optimum, and *Planning + complete Execution* ('Plan+compExec') schemes.

The *Planning + limited Execution* configuration of the proposed technique, that utilizes more frequently the opportunistic D2D-aided communication mode under lower densities of nodes (Figure 5) thanks to the implementation of the (limited) *real-time adaptation* step in the *Execution* phase, outperforms the *Planning only* configuration. Actually, it achieves a performance close to that obtained by the optimum scheme. Indeed, small differences, always below 8%, in the energy consumption with respect to SH traditional are observed between the *Planning + limited Execution* configuration and the optimum scheme. These differences reduce below 1% when $T_{max}$ is set to 30s and the number of nodes in the scenario is 1500 (Figure 6.c). This shows that the proposed technique combining the anticipated scheduling and communication mode selection decisions derived by the *Planning* phase, and the real-time decisions derived by the *real-time adaptation*, can achieve near-optimal performance.

Finally, the proposed technique introduces with the *Planning + complete Execution* configuration the potential to exploit the selected relays capabilities to further reduce the transmissions energy consumption. The selected relay is free to fully exploit the time available to complete the upload to find more efficient locations from where to perform the fragment transmission, or to use another relay to forward the fragment to the BS. These additional options to perform the transmissions significantly increase the possibility to find locations where the cellular link to the BS is under good (e.g. LOS conditions and shorter distances) communication conditions that require much lower energy consumption. The results reported in Figure 6 show that the *Planning + complete Execution* configuration outperforms all other schemes, including the optimum, that does not grant to the selected relay the freedom to decide where and when to upload the received fragment.

### 3) Communication and computation overheads

The proposed technique relies on the support from the BS to make some context information available to the mobile nodes in the cell: 1) local map of the region detailing streets and buildings; 2) maps of the nominal cellular and D2D path loss; and 3) density and distribution of nodes in the cell.

It has been estimated that the size of the transmitted local map is approximately 6 KB in scalable vector graphics format. To guarantee that 99.99% of the selected source nodes receive the map before the start of their transmissions (considering an scenario where nodes move in and out the cell, and where source nodes are selected randomly with the start of their transmissions following a Poisson distribution with a rate $\lambda_{req}$ = 1/10), the BS should transmit it every 10s. The cost of these broadcast downlink transmissions from the BS represents an increase of the energy consumption for the configurations of the proposed technique that ranges from 0.18% (*Planning only* configuration in the scenario with 100 nodes and $T_{max}$=10s) to 0.82% (*Planning+Complete Execution* configuration in the scenario with 1500 nodes and $T_{max}$=30s).

For the path loss map, we propose two different approaches considering their potential communication overheads. We consider the BS transmits the cellular path loss map of 24 KB[6] in the downlink using a broadcast channel every 10s to guarantee that 99.9% of the selected source nodes had received it. The cost of these transmissions would represent an increase in the energy cost for the proposed technique that ranges from 0.72% to 3.95%. The D2D path loss map of 19.2 KB[7] is transmitted on demand to the selected source nodes and relay nodes to avoid consuming excessive resources of the BS. In this case, the increase in the energy consumption for the configurations of the proposed technique ranges from 1.14% to 4.62%.

Finally, it is considered that the BS estimates the nodes density and distribution using the measurement reports they transmit. In addition, the BS can make this information available to the nodes in the cell using its broadcast channel at a negligible additional cost.

The proposed technique also utilizes standard mechanisms to discover and select the neighboring nodes when a D2D-aided transmission is executed. In particular, we consider the source nodes implement the Model B of the UE-to-Network relay discovery mechanism. This model is executed on demand when the source node needs to discover the neighboring nodes, and it is based on the exchange of 4 messages in the Physical Sidelink Discovery Channel (PSDCH): a *request* from the source node, multiple *responses* from the relays that receive the *request*, and a *direct communication request* and *direct communication accept* to establish the D2D link. The conducted evaluation shows that the energy cost of this discovery mechanism represents an increase for the proposed technique that ranges from 0.002% to 0.95%.

Regarding the computation overhead, and the associated CPU energy consumption associated to it, we provide, in the appendix, a conservative estimate of the number of floating point operations that need to be executed in each control interval, which results in values with an order of magnitude of $10^6$, and of the average additional power consumption, resulting in the order of 1mW. This additional consumption can be considered low if it is compared with the power consumption of typical devices CPUs even in the idle state (see the appendix).

This comprehensive analysis shows that the energy transmission gains reported in Section VIII.C.2) can be achieved at no significant communication and computational overheads of the proposed technique. An upper-bound estimate shows that the energy cost of the communication and computation overheads represent an increase of approximately 11% compared with SH traditional. This diminishes the energy gains achieved when only accounting for the transmission energy cost (that range from 60% to 90%), but it still shows the high potential of the efficient integration of NGO in cellular networks.

### IX. CONCLUSIONS

This paper has proposed a novel two-phase technique to facilitate the integration of NGO into D2D-aided cellular networks. Our solution builds on a new concept of graph that is utilized to represent all possible communication modes (including opportunistic cellular and opportunistic D2D-aided cellular) along the time available to complete the transmission. The graph utilizes geographic and link-context contextual information (e.g. stochastic models about the nodes location and distribution) to predict the evolution of the networking conditions and network connectivity, and to estimate the energy cost of each communication mode. These estimates are first utilized to perform an offline selection of the communication mode and transmission scheduling. These initial decisions are then re-visited, at execution time, using more up-do-date information about the network state to adapt them to the changes in the network state as the transmission progresses. The proposed technique has shown important energy efficiency benefits compared to traditional single-hop cellular communications. The proposed technique achieves its highest performance when the selected relays in the opportunistic D2D-aided cellular connections are granted the freedom to decide the communication mode and time instant to perform the transmission of the data they receive from the source node. In this case, the proposed technique can reduce the transmission energy consumption compared to single-hop cellular communications by more than 90%. This energy gains are achieved at the expense of a small transmission and computational overhead ($\sim 11\%$ energy cost increase).

---

[6] This size has been estimated considering that each value is encoded using 7 bits, and the size of the map is $O(L^2 − (B^2 · Bwidth^2))$, where L is the scenario size, $B$ is the number of buildings in scenario, and $Bwidth$ is the building plus street width.

[7] Each D2D path loss value is encoded using 8 bits, and the transmitted map only includes the D2D range deprived of the buildings and streets, i.e. the size of the map is $O((L^2 − (B^2 · Bwidth^2)) R_{D2D})$.

## Acknowledgment

This work has been partially funded by the Spanish Ministry of Science, Innovation and Universities, AEI, and FEDER funds (TEC2017-88612-R), the Ministry of Science, Innovation and Universities (IJC2018-036862-I), the UMH ('*Ayudas a la Investigación e Innovación de la Universidad Miguel Hernández de Elche 2018*'), and by the European Commission under the H2020 REPLICATE (691735), SoBigData (654024) and AUTOWARE (723909) projects.

## Appendix: complexity overhead

Considering the complexity overhead of the proposed solution, and the associated energy consumption, the major impact is given by the execution of the planning phase to compute the expected energy cost $\hat{C}_{D2D-aided*}(S; k)$ in (12), i.e., the expected cost of a D2D-aided transmission which exploiting a relay in the D2D coverage area of the source in each of the eligible time-slots (indexed by k). This computation dominates all the remaining contributions to the energy cost associated to the complexity overhead. Therefore, we will focus on this term to obtain a reliable conservative estimate of the additional energy consumption.

In input to the computation the following quantities are required:

- the costs $c_1(t_k), c_2(t_k), \ldots, c_Q(t_k)$. These costs are related to the nominal channel gains by simple path loss modeling functions. We assume that a map of these costs is available to each computing entity at the beginning of the process;

- the probabilities $p_{j(s,k)}(j)$, $\forall j \in \{j_{le}(s,k), \dots, j_{ue}(s,k)\}$, which are obtained from (8) and, in turn, from (7);

- the conditional probabilities $p_{C^*}\left(c_q^{(s,k)}|j\right)$, for each $j \in \{j_{le}(s,k), \dots, j_{ue}(s,k)\}$ and each $q \in \{1, \dots, Q\}$, which are obtained from (6);

- the values of $p_C(c_q)$, $\forall q \in \{1, \dots, Q\}$, which are directly obtained by the available density of nodes in the cell.

Let's start from the computation in (6). The value of $p_{C^*}\left(c_q^{(s,k)}|j\right)$ needs to be computed for each $j \in \{j_{le}(s,k), \dots, j_{ue}(s,k)\}$ and each $q \in \{1, \dots, Q\}$. However, it is easy to see that the operations required for performing the computation for $j = j_{ue}(s,k)$ already provide, as intermediate results, most of the values required to compute $p_{C^*}(c_q^{(s,k)}|j)$ for the smaller values of $j$. More specifically, the computation can be broken down as follows.

For $q = 1$ (first row of the right-hand side of eq (6))

#1 subtraction to compute $1 - p_C(c_q)$

#($j_{ue} - 1$) multiplications to compute $\left(1 - p_C(c_q)\right)^j$, $\forall j \in \{j_{le}(s,k), \dots, j_{ue}(s,k)\}$

#($j_{ue} - j_{le} + 1$) subtractions to compute $1 - \left(1 - p_C(c_q)\right)^j$, $\forall j \in \{j_{le}(s,k), \dots, j_{ue}(s,k)\}$.

For $q \in \{2, \dots, Q - 1\}$ (second row of the right-hand side of eq (6)):

($Q - 2$) sums to compute each of the ($Q - 1$) partial sums $\sum_{q'=1}^{q} p_C\left(c_{q'}\right)$, $\forall q \in \{2, \dots, Q - 1\}$

#($Q - 2$) subtractions to compute the terms $\left(1 - \sum_{q'=1}^{q} p_C\left(c_{q'}\right)\right)$, $\forall q \in \{2, \dots, Q - 1\}$

#($Q - 2$) * ($j_{ue} - 1$) multiplications to compute the terms $\left(1 - \sum_{q'=1}^{q} p_C\left(c_{q'}\right)\right)^j$, $\forall j \in \{j_{le}(s,k), \dots, j_{ue}(s,k)\}$

#($Q - 2$) subtractions to compute each of the partial sums $\sum_{q'=1}^{q-1} p_C\left(c_{q'}|j\right)$

#2 * ($Q - 2$) * ($j_{ue} - j_{le} + 1$) subtractions to compute the final sum $1 - \left(1 - \sum_{q'=1}^{q} p_C\left(c_{q'}\right)\right)^j - \sum_{q'=1}^{q-1} p_C\left(c_{q'}|j\right)$.

For $q = Q$ (third row of the right-hand side of eq (6)):

#2 subtractions to compute $1 - \sum_{q'=1}^{Q-1} p_{C^*}\left(c_{q'}|j\right)$. Note that the partial sum $\sum_{q'=1}^{Q-2} p_C\left(c_{q'}|j\right)$ is already available from the preceding iteration.

Overall, we have ($Q - 1$) * ($j_{ue} - 1$) multiplications and

$1 + (j_{ue} - j_{le} + 1) + (Q - 2) + (Q - 2) + (Q - 2) + 2 * (Q - 2) * (j_{ue} - j_{le} + 1)$
$= (Q - 2) * (2(j_{ue} - j_{le}) + 5) + (j_{ue} - j_{le}) + 2$ sums or subtractions.

To obtain the desired result in (12), additional $Q * (j_{ue} - j_{le} + 2)$ multiplications plus $Q * (j_{ue} - j_{le}) + Q - 1$ sums or subtractions are required.

Finally, the above described computations have to be repeated for each of the eligible time slots, let's say 30, indexed by $k$.

It is clear that the system parameter which mostly affects the computational overhead is the number of tiles, $Q$, eligible to be the possible possible location of potential relays. As an example, considering the *entire* surface, i.e., *without considering the techniques introduced to reduce the number of eligible tiles $Q$*, of a D2D circular coverage region with radius 100m, and tiles with side 5m, one would obtain values of $Q$ in the order of 1300 tiles. A similar order of magnitude for the number of tiles, *or even less*, could be obtained even with a more fine-grain tessellation, e.g., 1m-wide tiles (the one we used in the simulations), by excluding buildings and, e.g., considering only sidewalks. For instance, in our simulations, the average number of tiles per coverage region was 640, obtained with a coverage range of 80m.

Assuming $j_{ue}$ in the order of 10 (which is an upper bound on the value obtained in all our simulations), we can conservatively estimate a number of floating point operations in the order of few hundred thousand, i.e., less than 1 Million. In the following, we will consider a reasonable upper bound for it to be 1M ($10^6$) floating point operations. These operations need to be carried out at the beginning of the process, in the *planning* phase. In the *execution* phase, *with real-time adaptation*, the number of operations per time interval is much lower, since it takes into account the actual locations of the sole potential relays that have been found. However, even assuming that the number of floating point operation of the planning phase is executed in each slot, considering slots of duration one second, this corresponds to an average of (less then) $10^6$ floating point operations per second.

CPU architectures have a high variability in terms of computational energy efficiency. In [28], an energy per instruction figure of 11nJ is provided for a core duo processor. From other sources, e.g., [29], an estimate of ~1nJ per operation can be extrapolated as a reliable estimate of the order of magnitude of the energy cost of a single floating point operation in modern processors, with values roughly comprised in the interval [0.5, 2] nJ. With this figure, our conservative estimate of $10^6$ operations per second translates into an additional average power consumption of 1mW. A smartphone CPU, in the idle state, consumes an amount of power in the order of hundreds of mW, see, e.g., [30] Table I. Therefore, the average power consumption imposed by our algorithm represents less than 1% of the average power consumption in the idle state.